\documentclass[prb,twocolumn,showpacs]{revtex4}

\usepackage{graphicx}

\begin{document}

\title{Room-temperature manipulation and decoherence of a single spin in diamond}
\author{R. Hanson, O. Gywat and D. D. Awschalom}
\affiliation{Center for Spintronics and Quantum Computation, University of California, Santa Barbara, California 93106, USA}

\date{\today}

\begin{abstract}
We report on room-temperature coherent manipulation of the spin of a single nitrogen-vacancy center in diamond and a study of its coherence as a function of magnetic field. We use magnetic resonance to induce Rabi nutations, and apply a Hahn spin echo to remove the effect of low-frequency dephasing. A sharp rise in the decoherence rate is observed at magnetic fields where the nitrogen-vacancy center spin couples resonantly to substitutional nitrogen spins via the magnetic dipolar coupling. Finally, we find evidence that away from these energy resonances spin flips of nitrogen electrons are the main source of decoherence.
\end{abstract}



\maketitle

The study of single quantum systems is interesting both for testing fundamental laws of physics as well as for practical purposes, as computing with quantum systems promises an enormous increase in computing power and quantum communication allows secure information exchange~\cite{NielsenChuang}.
In the solid state, coherent control of single quantum systems has been achieved in a number of systems, e.g. superconducting Cooper pair boxes~\cite{Nakamura99} and electron spins in quantum dots~\cite{PettaScience05}. Among these, the nitrogen-vacancy (N-V) center in diamond~\cite{Davies76} is unique, because its spin exhibits a long coherence time that persists up to room-temperature~\cite{Kennedy03}, whereas most other systems only allow coherent control at cryogenic temperatures.

Coherent manipulation of N-V centers on large ensembles was first achieved many years ago~\cite{Glasbeek88,Charnock01}. Recently, however, coherent rotations and spin echoes of a single N-V center spin were reported by Jelezko et al.~\cite{JelezkoRabiPRL04}. This landmark experiment, that has not been reproduced by a different group thus far, demonstrates that the N-V center provides a testbed for quantum manipulation in the solid state at room temperature~\cite{Jelezko2qubitPRL2004,PopaPRB04}. On the other hand, single-center spectroscopy allows the study of the local environment of the N-V center~\cite{GruberScience97} and has already unveiled anisotropic spin interactions and magnetic dipolar coupling to spins of other defects in diamond~\cite{EpsteinNatPhys05}. Recent results of these studies include the observation of strong coupling between a single N-V center and the spin of a single substitutional nitrogen atom~\cite{Gaebel06,HansonPRL06} and the measurement of the spin relaxation time of a single nitrogen electron spin~\cite{HansonPRL06}. By combining single-center spectroscopy with coherent control, the coherent interaction of the N-V center spin with its environment can be probed, which might ultimately lead to coherent quantum circuits~\cite{Jelezko2qubitPRL2004}.

Here, we report coherent control of the electron spin state of a single N-V center at room temperature. We demonstrate that we can coherently drive single-spin rotations (Rabi nutations) and undo low-frequency dephasing by application of a Hahn spin echo sequence. We use this capability to study the coherence of the N-V center as a function of applied magnetic field. By comparing the magnetic-field dependences of the decoherence rate and the photoluminescence, we find that the coherence of the N-V center spin is strongly affected by the resonant spin exchange with the surrounding nitrogen spins at specific magnetic fields. At other fields, we find evidence that the dephasing caused by spin flips of the nitrogen electrons is the main contribution to the observed decoherence. 

\begin{figure}[bp]
	\includegraphics[width=3.4in]{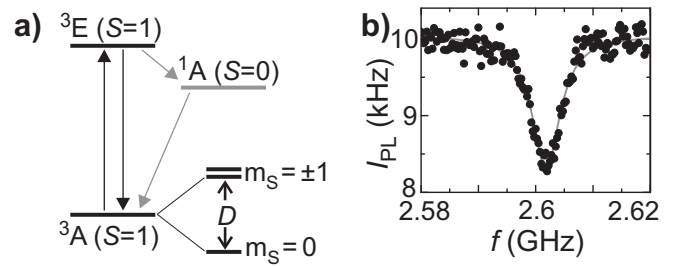}
	\caption{\textbf{\textbf{(a)}} Electronic level structure of the N-V center. The excited $^3E$ state is 1.945~eV above the $^3A$ ground state. The spin sublevels of the electronic ground state are manipulated in this work. \textbf{(b)} Electron spin resonance (ESR) between the $m_S\!=\!0$ and $m_S\!=\!-1$ spin levels of NV 14 at $B$~=~100~G.}
	\label{fig:fig1}
\end{figure}

The N-V center consists of a substitutional nitrogen atom next to a vacancy in the diamond lattice. We study the negatively charged state of the N-V center (N-V$^-$), which has a level structure as depicted in Fig.~\ref{fig:fig1}(a). It has a spin triplet ($^3$A) ground state, with a zero-field splitting $D$= 2.88~GHz between the sublevels with spin z-component $m_S\!=\!0$ and $m_S\!=\!\pm1$~\cite{GroundStates}. The spin is quantized along the N-V symmetry axis, a $<\!\! 111 \!\! >$ crystal axis~\cite{Loubser78}. There is a strong optical transition to an excited ($^3$E) triplet state that conserves spin. Linearly polarized optical excitation preferentially pumps the spin system into the ground state $m_S\!=\!0$ level~\cite{Harrison2004}, enabling efficient optical initialization of the spin state. Also, the average photon emission rate is substantially smaller for transitions involving the $m_S\!=\!\pm 1$ levels than for the $m_S\!=\!0$ level~\cite{JelezkoAPL02}, which allows readout of the spin state by the photoluminescence intensity $I_{PL}$. These two effects have been attributed to spin-dependent intersystem crossing from the $^3$E state to a nearby singlet ($^1$A) state~\cite{Loubser78,Manson06,Nizovstev03}.

We study five different single N-V centers in a single-crystal high-temperature high-pressure (type Ib) diamond, commercially available from Sumitomo Electric Industries. This diamond contains nitrogen (N) impurities with a density of $10^{19}-10^{20} $~cm$^{-3}$. These impurities (also known as P1 centers) contain one unpaired electron carrying a spin of $\frac{1}{2}$. 
The N-V centers are typically spaced by a few $\mu$m, much larger than the spatial resolution of our setup ($\approx 0.3~\mu$m), allowing the study of single N-V centers. We detect the phonon-broadened $^3$E-$^3$A transition of the N-V center using non-resonant photoluminescence microscopy along the [001] crystal axis (for details on the setup see Ref.~\onlinecite{EpsteinNatPhys05}). We identify single N-V centers through photon antibunching measurements and Electron Spin Resonance (ESR) measurements. All experiments are performed at room temperature.

We precisely align the external magnetic field $B$ with the [111] crystal axis, which is the symmetry axis for all centers studied in this work. We will refer to this axis as the $z$-axis.
The spin Hamiltonian of the N-V center $H_{NV}$ can then be written as 
\begin{equation}
	H_{NV}=D S_{z}^2 + g \mu_B B S_{z} + \vec{S}\ \bar{A}\ \vec{I},
\label{Eq:Hamiltonian}
\end{equation}
where $\mu_B$ is the Bohr magneton, $\vec{S}$ is the N-V center electron spin operator, $\bar{A}$ is the hyperfine tensor~\cite{Loubser78,Charnock01} and $\vec{I}$ is the spin operator of the N-V center nitrogen nucleus. Virtually all nitrogen atoms in our diamond are $^{14}$N with total nuclear spin $I$=1. The electron $g$-factor is 2.00.

\begin{figure}[!htb]
	\includegraphics[width=3.4in]{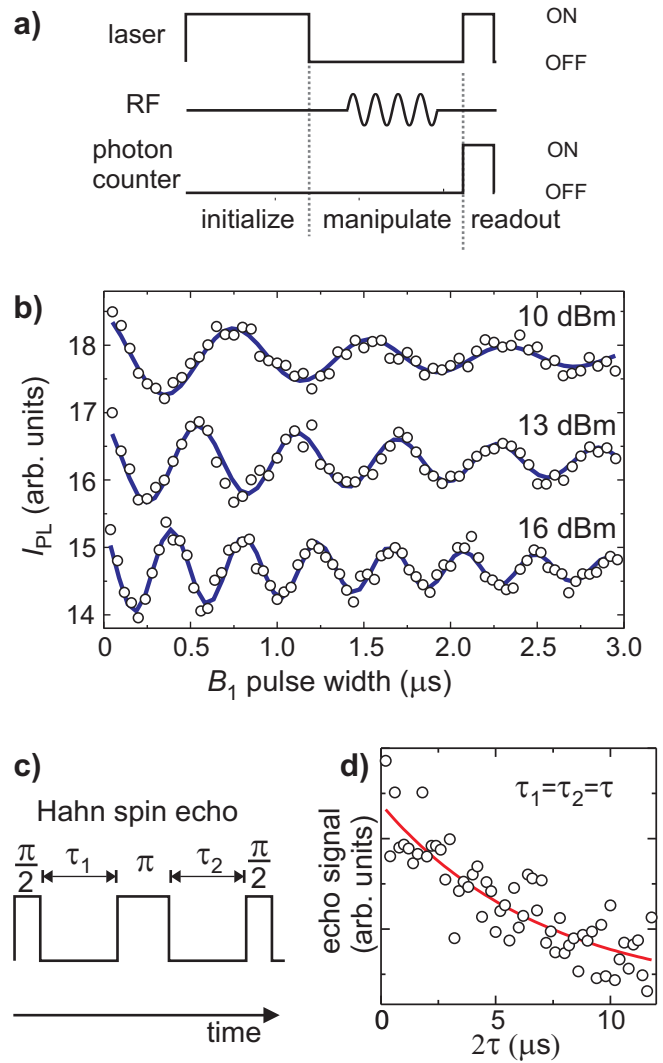}
	\caption{Coherent manipulation of the electron spin of a single N-V center. \textbf{\textbf{(a)}} Three-step sequence applied: the first laser pulse (typically 5~$\mu$s long) initializes the spin, then the spin manipulation is carried out in the dark, and finally the spin state is read out by a second laser pulse (typically 2~$\mu$s long)~\cite{PulseSeq}. The sequence is typically repeated 1000 times to increase the signal. Laser power is 500~$\mu$W.
\textbf{(b)} Driven coherent oscillations (Rabi nutations) of NV14 at $B$=850~G. The spin is rotated between the $m_S\!=\!0$ and the $m_S\!=\!-1$ level, leading to alternating high and low $I_{PL}$. The Rabi nutation frequency is proportional to the square-root of the applied RF power, as expected.
Curves are offset for clarity. \textbf{(c)} Hahn spin echo pulse sequence. \textbf{(d)} Hahn spin echo signal of NV14 at $B$=850~G as a function of the total delay $\tau_1+\tau_2$.}
	\label{fig:fig2}
\end{figure}

We use magnetic resonance to manipulate the N-V center electron spin state. An AC magnetic field, perpendicular to the static magnetic field, is generated by sending a radio-frequency (RF) alternating current through a gold wire of 25~$\mu$m diameter that runs close to the N-V center. This flexible approach allows us to study many different centers in a single sample. For a wire-to-N-V center distance of 20~$\mu$m, the AC magnetic field at the site of the N-V center is about 1~G for 16~dBm output power at the RF source. Figure~\ref{fig:fig1}(b) shows the typical continuous wave (CW) ESR signal for a single center (in this case NV14). Away from the resonance, the photoluminescence is high because the spin is strongly polarized into the $m_S\!=\!0$ state by the continuous optical excitation. When the applied frequency matches the energy splitting between the $m_S\!=\!0$ and $m_S\!=\!-1$ levels, the AC magnetic field induces rotations between the spin levels. This reduces the spin polarization and therefore the measured photoluminescence intensity $I_{PL}$ drops.

For coherent control of the spin state, we separate the polarization, the spin manipulation and the readout in time by applying the sequence depicted in Fig.~\ref{fig:fig2}(a). First, a strong laser pulse initializes the spin in the $m_S\!=\!0$ state. Then the laser is turned off, and the AC magnetic field is pulsed to coherently manipulate the spin state. Finally, the spin state is read out by a second laser pulse. This way, we avoid the decoherence that can be induced by the optical excitation~\cite{JelezkoRabiPRL04}.

Upon application of the AC magnetic field, the spin is coherently driven between the two spin sublevels. The probability $P_{m_S=0}$ to find the N-V center in the $m_S\!=\!0$ state thus varies periodically, as described by Rabi's formula~\cite{Sakurai}
\begin{equation}
	P_{m_S=0}\! =1\!-\frac{f_{1}^2}{f_{1}^2\!+\! \Delta f^2}\ {\rm sin}^2\ [\pi \sqrt{f_1^2\!+\!\Delta f^2}\ t],
\label{Eq:Rabi}
\end{equation}
where $f_1$ is the Rabi nutation frequency, $\Delta f$ the detuning from the resonance frequency and $t$ the time from the start of the AC magnetic field pulse. The Rabi frequency depends on the amplitude $B_1$ of the AC magnetic field: $f_1= \frac{1}{2} g \mu_B B_1/h$, where $h$ is Planck's constant and the factor of $\frac{1}{2}$ is due to the rotating wave approximation.

In Fig.~\ref{fig:fig2}(b) we plot $I_{PL}$ as a function of pulse width for three different RF output powers. We observe that $I_{PL}$, which is proportional to $P_{m_S\!=\!0}$, oscillates with a frequency that depends on the applied power. The blue curves are fits to the data using a cosine multiplied by an exponential with a single decay constant $T'_2$ (note that this time is different from the dephasing time $T^*_2$~\cite{LievenRMP}). These measurements demonstrate our ability to coherently manipulate the spin state of a single N-V center.

That each Rabi oscillation curve is well fit by a single-frequency cosine is surprising at first sight, since the hyperfine coupling to the nitrogen nucleus of the N-V center is expected to split the resonance condition into three frequencies, spaced by about 2~MHz~\cite{Loubser78} (see Eq.~\ref{Eq:Hamiltonian}). From Eq.~\ref{Eq:Rabi} we see that, for a fixed value of $B_1$, the nutation frequency depends on the spin state of the nitrogen nucleus. As we average over different nuclear spin configurations, a beating pattern is expected in Fig.~\ref{fig:fig2}(b). The absence of this beating could be explained by a hyperfine-induced polarization of the nitrogen nuclear spin under strong optical excitation. Such a polarization was observed recently around 500~G~\cite{Gaebel06}, but is possible at other fields as well. Indeed, we observe a single Rabi frequency at all magnetic fields probed (50-1020~G). Note also that ``forbidden'' transitions, in which the nuclear spin is simultaneously changed with the electron spin, are not observed here.

For long pulses, the oscillations damp out due to interactions with the environment. We note that the decay time $T'_2$ increases with increasing Rabi frequency, which can be ascribed to the refocussing effect of the Rabi nutations: continuous driving can be viewed as a series of concatenated $\pi$-pulses~\cite{LievenRMP}. The data in Fig.~\ref{fig:fig2}(b) therefore suggests that some of the observed damping is due to dephasing and thus can be eliminated. To test this hypothesis, we apply a Hahn spin echo to the N-V center. The Hahn pulse sequence is depicted in Fig.~\ref{fig:fig2}(c). A $\pi/2$ pulse creates a coherent superposition of the $m_S\!=\!0$ and $m_S\!=\!-1$ states, which is allowed to dephase during a time $\tau_1$. After a $\pi$-pulse, the dynamics are reversed and a spin echo occurs after a time $\tau_2$ equal to $\tau_1$. We apply a final $\pi/2$ pulse to map the echo signal to the readout basis. By fixing $\tau_1$ and measuring the spin echo signal as a function of $\tau_2$, we have confirmed that the echo signal is maximum for $\tau_1=\tau_2$ (data not shown).

Figure~\ref{fig:fig2}(d) shows the echo signal as a function of $2 \tau$ (here $\tau_1=\tau_2=\tau$). The echo signal decays exponentially on a timescale of $T_2$=6~$\mu$s, about three times longer than the decay time of the Rabi nutations. This demonstrates that for this N-V center the fluctuations in the environment that dominate the decay are slow on the microsecond timescale. The influence of higher-frequency dephasing sources can in principle be eliminated by applying multiple spin echoes or more elaborate pulse sequences.

\begin{figure}[!htb]
	\includegraphics[width=3.4in]{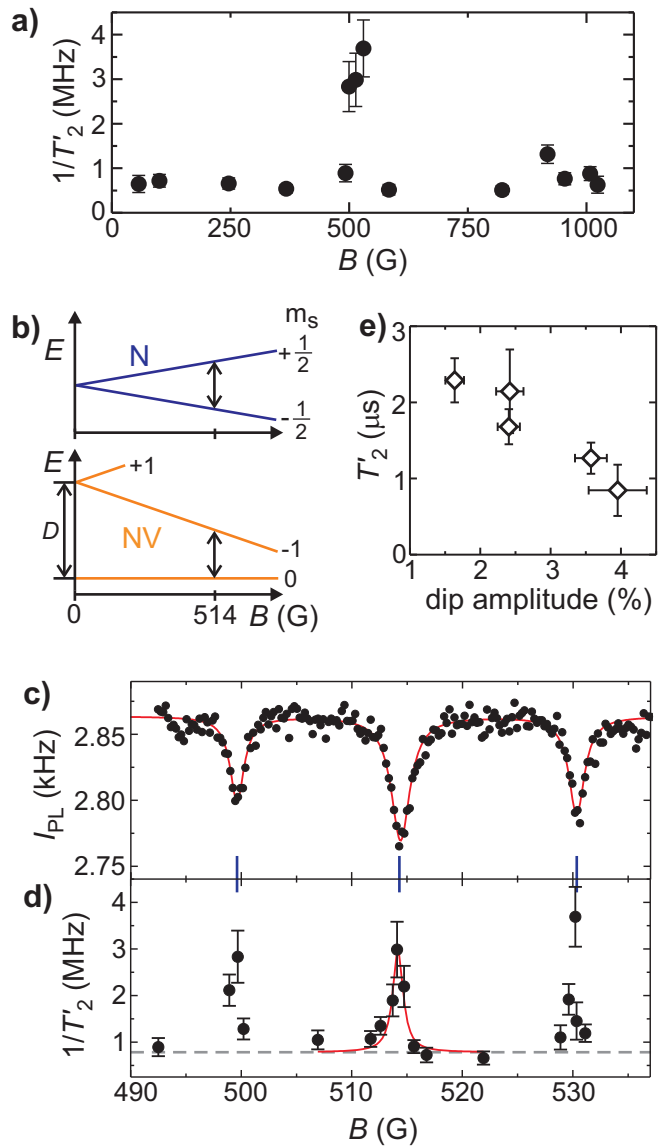}
	\caption{Dependence of the coherence on the applied magnetic field $B$. \textbf{(a)} Decoherence rate 1/$T'_2$ of NV14 as a function of $B$. \textbf{(b)} Energy levels of the N-V center and of N electron spin as a function of $B$ along the N-V symmetry axis. An energy resonance occurs around $B$=514~G. \textbf{(c)} $I_{PL}$ of NV14 as a function of $B$. The dips reflect the reduced spin polarization of the N-V center due to resonant spin exchange with the N electron spin through magnetic dipolar interaction~\cite{EpsteinNatPhys05,HansonPRL06}. \textbf{(d)}1/$T'_2$ of NV14 as a function of magnetic field, showing stronger decoherence at the fields where $I_{PL}$ dips. The dashed line depicts the average off-resonance value of 1/$T'_2$. The red lines in (c) and (d) are Lorentzian fits. \textbf{(e)} $T'_2$ at $B$=850~G versus normalized amplitude of the dip in $I_{PL}$ for five different N-V centers.}
\label{fig:fig3}
\end{figure}

We investigate the sources of decoherence in more detail by measuring Rabi nutations over a wide range of magnetic field. In Fig.~\ref{fig:fig3}(a) we plot $1/T'_2$ as a function of $B$. We observe that $1/T'_2$ is more or less constant over the whole range, except around $B$=514~G. 
As was found in previous work~\cite{vanOortPRB90,EpsteinNatPhys05,HansonPRL06}, this magnetic field marks an energy resonance between the $m_S=0$ and $m_S=-1$ levels of the N-V center and the $m_S=+1/2$ and $m_S=-1/2$ electron spin levels of substitutional N impurities, see Fig.~\ref{fig:fig3}(b). The magnetic dipolar coupling leads to resonant spin exchange at this field, and this could be the cause of the observed reduction of coherence.

In order to find evidence for this hypothesis, we compare the magnetic field dependences of 1/$T'_2$ and $I_{PL}$. At the resonance condition, the magnetic dipolar coupling leads to spin exchange between the N-V center and the N spins, and the resulting reduction in spin polarization of the N-V center translates into a dip in $I_{PL}$ (see also Refs.~\onlinecite{EpsteinNatPhys05,HansonPRL06}). As can be seen in Fig.~\ref{fig:fig3}(c), the resonance has sidepeaks due to the strong hyperfine interaction of the N impurity~\cite{SmithPR59,EpsteinNatPhys05}. Fig.~\ref{fig:fig3}(d) shows 1/$T'_2$ over the same magnetic field range. Peaks in 1/$T'_2$ are observed at exactly the same magnetic fields where $I_{PL}$ dips, which demonstrates that the magnetic dipolar coupling is indeed the cause of the reduced coherence.

Both the central peak in 1/$T'_2$ and the central dip in $I_{PL}$ are well fit to a Lorentzian (red lines in Figs.~\ref{fig:fig3}(c)-(d)). The width of the dip in $I_{PL}$ is a bit larger than the width of the peak in 1/$T'_2$, which is likely due to broadening induced by the continuous laser excitation. Note that this broadening is absent in the measurement of 1/$T'_2$, since the laser is turned off during the spin manipulation.

Away from the energy resonances, the magnetic dipolar coupling becomes inefficient in exchanging spins and 1/$T'_2$ goes down. However, the N spins can still contribute to decay of the Rabi nutations via the $z$-component of the magnetic dipolar field that they create at the site of the N-V center. If an N electron flips its spin, this magnetic field is reversed. Such events shift the total magnetic field felt by the N-V center and lead to dephasing.
The more N spins are nearby, the more fluctuations in the magnetic field and the stronger the decay. Therefore we expect a relation between $T'_2$ and the strength of the dipolar fields if the N spins are also the dominant source of decoherence away from the resonances. We use the amplitude of the dips in $I_{PL}$ at $B$=514~G as a measure for the strength of the dipolar fields at a given N-V center. In Fig.~\ref{fig:fig3}(e) we plot $T'_2$ versus the dip amplitude normalized to the off-resonance $I_{PL}$. The five centers investigated follow the expected trend, suggesting that dephasing due to N electron spin flips is the main decoherence mechanism away from the energy resonances.

Our results on single centers confirm and extend the findings from previous measurements on large ensembles of N-V centers. Hiromitsu and coworkers observed a rise in the Hahn echo decay rate around 514~G~\cite{Hiromitsu92}. Kennedy and coworkers found that the Hahn echo decay time was shorter for samples with a higher density of N impurities~\cite{Kennedy03}. Our results show that the local density of N spins determines the coherence time of single centers in our sample, with corresponding fluctuations observed from center to center.

In summary, we have demonstrated the ability to coherently manipulate the spin state of a single N-V center in diamond and probe its coherence. Using this tool, we have investigated the dependence of the coherence on magnetic field, showing that the magnetic dipolar coupling to substitutional N impurities is the dominant source of decoherence in this diamond. As a next step, coherent two-spin operations can be envisioned for N-V centers that are strongly coupled to a single N spin~\cite{Gaebel06,HansonPRL06}.

We thank  J.M. Elzerman, R.J. Epstein and F.M. Mendoza for discussions.
This work was supported by AFOSR, DARPA/MARCO, DARPA/CNID and ARO.


\begin{thebibliography}{99}
\bibitem{PulseSeq}
Multiple synchronized outputs of a pattern generator are used to control the laser via an Acoustic-Optical modulator, to gate the photon counter and to trigger an arbitrary waveform generator that in turn gates the high-frequency source leading to the desired sequence of RF bursts.
\end{thebibliography}
\end{document}